\documentclass[preprint,showpacs,preprintnumbers,amsmath,amssymb]{revtex4}

\usepackage{graphicx}

\usepackage{dcolumn}
\usepackage{bm}
\usepackage{color}

\begin{document}

\title{Point Contact Spectroscopy of Nb$_3$Sn Crystals: Evidence of a
CDW Gap Related to the Martensitic Transition.}

\author{R. Escudero}\email[Author to whom correspondence should be
addressed. Email address:]
{escu@servidor.unam.mx} 
\author{ F. Morales}
\affiliation{Instituto de Investigaciones en Materiales, Universidad
Nacional Aut\'{o}noma de M\'{e}xico. A. Postal 70-360. M\'{e}xico, D.F.
04510 MEXICO.}

\date{\today}

\begin{abstract}Two Single crystals of Nb$_3$Sn presenting the
martensitic anomaly at different temperature and shape, as observed
with specific heat measurements, were used to study structural
features in the electronic density of states with point contact
spectroscopy. At high temperature below the martensitic anomaly, we
observed different spectroscopic characteristics. One sample
displaying a well marked specific heat peak, shows a clear defined
structure in the differential conductance that evolves with
temperature and may be associated with changes on the density of
states due to the opening of a charge density wave gap. Those
features are very depending on the crystallographics characteristics
 of the single crystal examined.

\end{abstract}

\pacs{74.25-q Superconducting intermetallic alloys, 74.25.Bt Point contact spectroscopy}

\maketitle

\section{Introduction}
Nb$_3$Sn is a well known intermetallic compound with the cubic A15
crystal structure. This intermetallic alloy presents a cubic to
tetragonal martensitic transition at about 40-50 K, depending on the
crystal studied, and a superconducting transition at around 18 K
\cite{mai1,mailefert,Weger}. From the thermodynamic view point, a
martensitic is considered to be a first order transition. However,
particularly in this compound controversial results about the
thermodynamic order are not completely clarified \cite{vieland1, chu,
vieland2, labbe}. Quite recently it was established that
this transition is a second or weakly first order thermodynamic
transition \cite{escu}. One aspect related to this behavior is that
in many studied Nb$_3$Sn samples, frequently the martensitic anomaly
was observed only as a small feature, which may be related to diverse
causes; imperfections in the samples studied, impurities, defects,
vacancies, accumulated stress in the crystal structure at the moment
of growth process. Those different factors in the specimens may be
one of the reasons of the changes in the size and temperature of the
anomaly. In a recent paper \cite{escu} this anomaly was observed with
extraordinarily clarity in some crystals, but was very faint in
others.

The aim of this work is to study with point contact spectroscopy
(PCS) using some of the same single crystal specimens already
reported \cite{escu}, changes in the electronic density of states
when the high temperature anomaly is shown in different size and
temperatures \cite{weber}.

Accordingly, we used two single crystals that present the martensitic
anomaly, as observed by specific heat measurements, in different form
and temperature. Those crystal were well characterized by X-ray, and
present clear structural differences. In Fig.1 we show the lattice
dimensions of one of the crystals, and in  Table 1 and  Fig. 2 we
display the structural characteristics and the specific heat features
for  the  two used specimens. Our PCS results point to differences on
the electronic density of states as observed in the differential
conductance versus bias voltage of the two studied specimens. Our
initial assumption about this behavior is that the martensitic
anomaly, could be related to a Peierls distortion via dimerization of
Nb atoms and nesting at the Fermi surface, which in turn opens a
Charge Density Wave (CDW) gap, as was  formulated by Gor'kov, Bhatt
and McMillan, and Bilbro and McMillan, many years ago \cite{gorkov73,bhatt76}. Accordingly to
those  experimental results we rise the hypothesis that the specific
heat anomaly will be sharper and well formed if no Nb vacancies
exist, thus  chains are completely formed; so deficiencies in the
chains will disrupt the Nb dimerization and produce an imperfect
nesting  of the Fermi surface giving an ill form of the CDW gap.

\begin{figure}[btp]
\begin{center}
\includegraphics[scale=0.6]{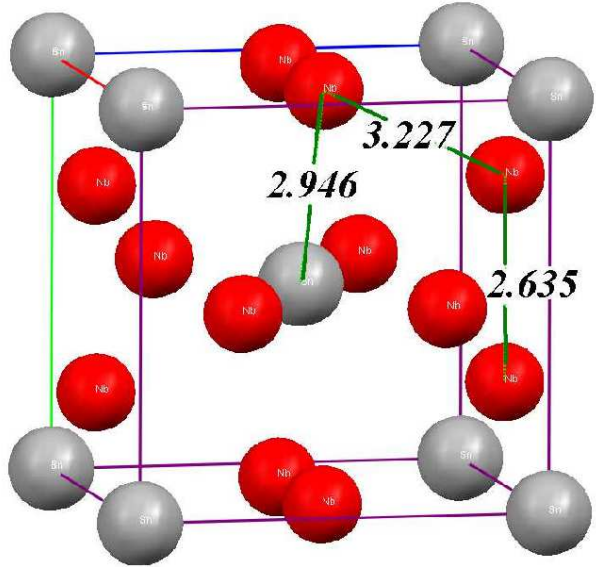}
\end{center}
\caption{(Color online) Crystalline structure
of Nb$_3$Sn, NS1 sample and lattice parameters as determined and given in Table 1. Distances
are  in  \AA, The grey (big) and red (small) spheres correspond
to Sn and Nb respectively, more details are given in Table 1.}
\label{fig1}
\end{figure}

\section{Experimental Details}

The study was performed in high quality single crystal samples. Many
of the details of the crystal characteristics can be found in ref
\cite{escu}. The crystals were growth by closed-tube vapor transport
with iodine vapor as the transport agent in a period of about four
months. Specific heat capacity measurements were performed from room
temperature to 2 K, under a magnetic field below 0.1 Oe.
For those measurements we used a thermal relaxation technique with a
Quantum Design calorimeter, and a Bachmann´s approach for determining
the specific heat capacity, as suggested by Lashley, et al
\cite{lash}. The crystals were measured and characterized by
magnetization versus temperature, resistance versus temperature, and
specific heat versus temperature. We present only specific heat
measurements which are the most typical results for two different
Nb$_3$Sn samples; named NS1, and NS4.

Point contact spectroscopy junctions were prepared using a fine tip
of a gold-tungsten  wire of 5 micron diameter. Measurements were
performed by contacting the Nb$_3$Sn samples with the 5 microns wire.
The point contact spectra were obtained using the standard modulation
technique consisting of  a resistance bridge  and lock-in amplifier.
More than 20 junctions were measured, here  we will present the most
typical results for two different Nb$_3$Sn samples. PCS measurements
show changes in  the differential conductance (dI/dV) versus bias
voltage (V) at different temperatures from high temperature to about
2 K. Data for sample NS4 were recorded from 80 K to below, whereas
for sample NS1 was from 60 K to below. Below the onset of the
martensitic anomaly, as specific heat data shows, we observed
structure in dI/dV-V in sample NS1, but non in sample NS4. At about
18 K, in both specimens we observed the formation of a single feature
related to the superconducting energy gap. The size values, as
determined at low temperature,  fit well with early tunneling studies
in this compound\cite{moore,geerk, shen}.

\section{Results and Discussion}

\subsection{Crystallographic characterization}

An important antecedent of the high quality of the crystals is that
others of the same batch, were used to perform  de Haas van Alphen
studies of the Fermi surface. It is important to mention that in
order to observe de Hass van Alphen oscillations, the  studies
require high quality single crystals with a minimum
number of imperfections or defects. \cite{arko78}.

For the crystallographic characterization we studied two samples by
X-ray diffraction, hereafter called NS1 and NS4, with mass about 5.31
and 9.5 mg, respectively. X-ray characteristics were taken at a
temperature of 298(1)K. In addition, we remark that no secondary
diffraction patterns were observed for possible impurities, neither
diffuse scattering. For each sample, a complete diffraction sphere
was collected \cite{siemens} at the highest available resolution
(0.62 \AA, 2$\theta_{max}$ = 70$^\circ$). A characteristic parameter
in a crystal is the  high value of the extinction coefficient, which
converges to identical values for both samples: so in those crystals
we found  0.82(18) for NS1 and 0.8(2) for NS4 \cite{sheldrick}.
Assuming that applied correction covers mixed primary and secondary
extinctions, this result suggests that samples should have similar
block sizes and similar concentrations of randomly distributed
dislocations \cite{masimov}.

\begin{table}
\caption{Crystallographic data for two single crystals NS1 and NS4.}
\begin{tabular}{lll}
  \toprule
  Compound & NS1 (5.31 mg) & NS4 (9.5 mg) \\    \colrule
    \botrule
  \textbf{Geometric Parameters}& &\\ \hline
  Distance& & \\ \hline
  Nb-Sn (\AA)  & 2.9460 (5) & 2.9366 (7) \\
Nb-Nb (\AA)  & 2.6350 (5)  &2.6266 (6) \\
Nb...Nb (\AA) &3.2272 (5)  &3.2169 (8) \\ \botrule
\end{tabular}
\label{table1}
\end{table}

In Table I we present some of the characteristics of the two
specimens measured. In Fig. \ref{fig1} is shown the crystal structure
of NS1 indicating the distances between atoms. Both samples are
characterized by rather short unit cell parameters, \emph{a} =
5.2700(9) and \emph{a} = 5.2531(13) \AA, for NS1 and NS4,
respectively, while the accepted value found in the literature for
crystalline Nb$_3$Sn is \emph{a} = 5.29 \AA\ \cite{pdf}.
Interestingly, NS1 and NS4 have significantly different cell
parameters, and, as a consequence, cell volume is reduced by
\emph{ca} 1\% in NS4, compared to NS1. Calculated densities present
the same 1\% drop. However, using diffraction data, a confident
interpretation of such a cell contraction in terms of intrinsic
vacancies in the alloy cannot be carried out, at least if departures
from Nb$_3$Sn stoichiometry remain small. In contrast, the high
resolution of diffraction data allows to accurately determine
distances in the crystalline structure. The shortest Nb...Nb
separation is reduced from 2.6350(5) \AA\ in NS1 to 2.6266(6) \AA\ in
NS4. In the same way, Nb...Sn separations in NS1 and NS4 are
2.9460(5) and 2.9366(7) \AA, respectively. These differences between
the two studied samples may be related with the presence of a high
number of defects in the NS4 sample.

\subsection{Specific heat measurements}

\begin{figure}[btp]
\begin{center}
\includegraphics[scale=1.2]{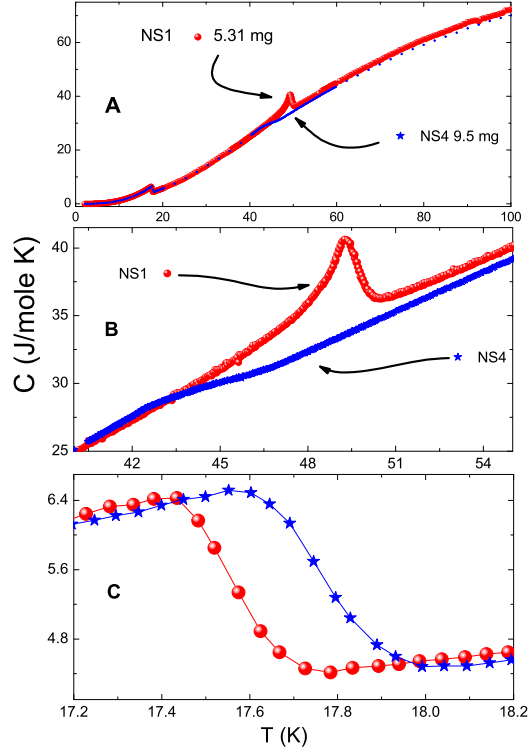}
\end{center}
\caption{(Color online)  Specific heat around the martensitic and superconducting
transitions of two studied samples. Note the different size of the martensitic anomaly
that is presented at  different temperatures for each
specimen, panel B. Panel C shows the superconducting onset differences in temperature, which is small
 but discernible.}
\label{fig2}
\end{figure}

Figure \ref{fig2} shows the specific heat of the two studied samples,
in the temperature range from 100 K to 2 K (panel A). There, the
martensitic transition and the superconducting transition are
exhibited. Panel B shows a zoom around the high temperature feature,
where a peak is clearly observed for sample NS1, whereas  sample NS4
shows only a small and  smeared feature. In this same panel it seems
that the onset temperatures are quite different. For sample NS1 the
onset temperature is about 50 K, whereas for sample NS4 is around
46.5 K, but difficult to clearly distinguish. In comparison panel C
shows the two superconducting transitions, In which the onsets are
well distinguished. An important precision is worth to make; In
comparing the onset temperatures in panel B and C, we observed that
sample NS1 has a martensitic transition temperature higher than
sample NS4, whereas  the superconducting transition in the same NS1
sample is lower than of sample NS4. One possibility of this distinct
behavior could be related to a decreasing of the electronic density
of states at the Fermi level because the electron population used to
form the CDW gap, this behavior was analyzed and
discussed in \cite{escu}.

\subsection{Point Contact Spectroscopic Characteristics}

In PCS, one very  important parameter is to know the relationship
between the electronic mean free path ($l$) and the point contact
diameter ($d$). For the determination of the point contact region of
electron injection,  it is assumed in general that the contact is
circular; three regimes can be identified and occur for the following
conditions: the ballistic regime when $l_e\gg d$, the thermal regime
when $l_e,l_{in}\ll d$, and the diffusive regime when $l_e\ll d\ll
\sqrt{l_{in}l_e}$. Where $l_e$, and $l_{in}$ are the elastic and
inelastic mean free paths, respectively. To extract the correct
spectroscopic information from the point contact spectrum, it is
necessary to determine the regime of the contact. In the thermal
regime there are heating effects in the point contact region, this
increases locally the temperature via Joule heating, and the
(dI/dV)-V characteristic resembles the resistance temperature
dependence of the sample. In this thermal regime the (dI/dV)-V curve
does not give information about the superconducting gap. Of course
the most appropriate regime to study spectroscopic features on the
DOS is the ballistic regime in first place, and the diffusive limit
as second option. However, experimentally one can find limitations to
work in the ballistic regime due to the size of the electronic mean
free path. So the diffusive limit is the second option where still we
can observe spectroscopic features as the energy gap or phonon
influence on DOS. In our experimental PCS junctions we used the
formula
 in  \cite{PCS} for the determination of $d$,
\begin{equation}
d=\frac{d\rho (T)/dT}{dR_{pc}(T)/dT}.
\end{equation}
In this equation  $\rho$ is the resistivity and $R_{pc}$ is the
differential resistance of the point contact at zero bias. We
estimate that $d$ is about 100 nm. Nb$_3$Sn is a type II
superconductor and has a short electronic mean free path of about 5
nm, this value is lower than $d$, thus  our point contacts will be in
the diffusive regime.

\begin{figure}[btp]
\begin{center}
\includegraphics[scale=1]{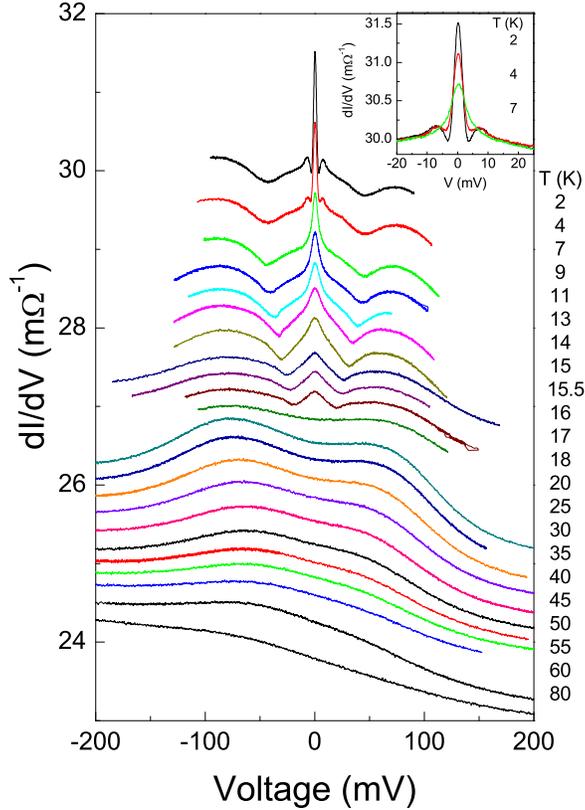}
\end{center}
\caption{(Color online) Differential conductance of Nb$_3$Sn(NS1)-Au
junction, which presumable has complete Nb chains. Note structure
above the superconducting temperature that may be related to the
opening of a CDW gap at the martensitic transition, assuming that
this is  generated by a Peierls distortion, according to Gor'kov,
Bhatt-McMillan, Bilbro-McMillan model. For clarity, the curves were displaced related
to the curve measured at 2 K. The inset shows three typical conductance
curves of the evolution of the superconducting
feature of the PCS junction, at 7, 4, and 2 K.}
\label{fig3}
\end{figure}

\begin{figure}[btp]
\begin{center}
\includegraphics[scale=1]{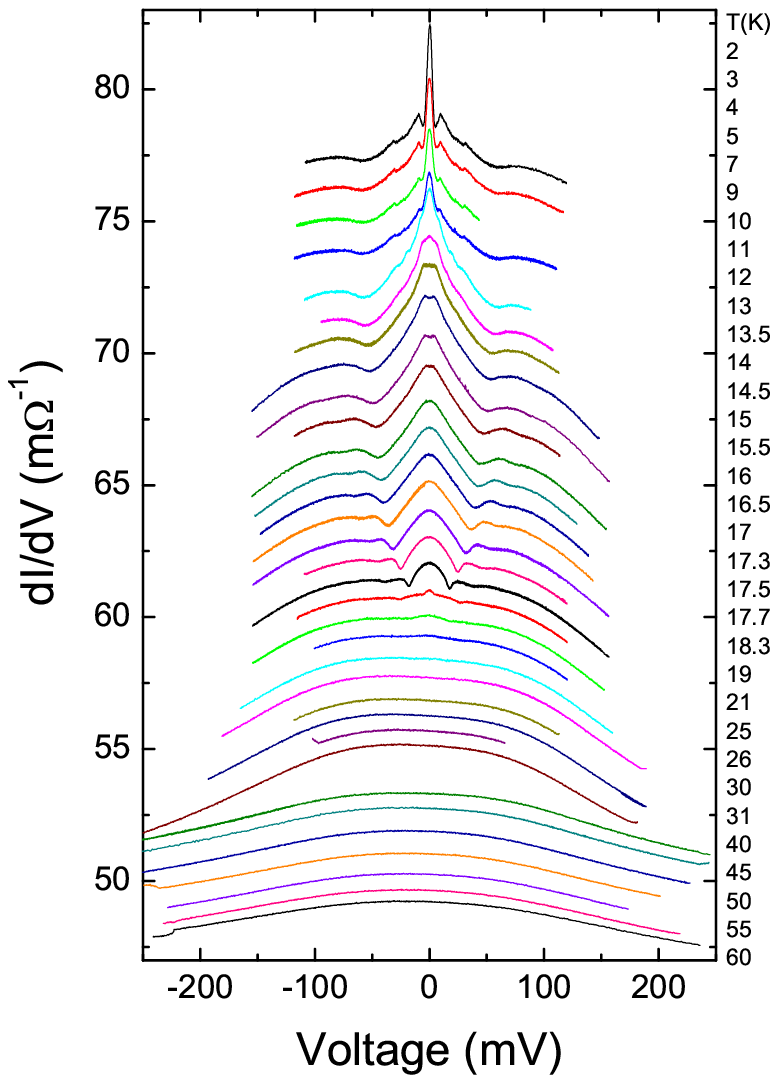}
\end{center}
\caption{(Color online) Differential conductance as a function of
bias voltage of a Nb$_3$Sn (NS4)-Au point contact junction at various
temperatures. Note the absence of structure for above 18 K. The curves
were displaced for clarity, related to the curve measured at 2 K.}
\label{fig4}
\end{figure}

Figs. \ref{fig3}, and \ref{fig4} show the differential conductance as
a function of the bias voltage measured at different temperatures
from 2 K to 80 and 60 K for two different point contact,
Nb$_3$Sn-AuW,  with samples NS1 and NS4 respectively.

At low temperature the curves show the typical feature of the
superconducting energy gap around zero bias. The feature clearly
evolves  as the temperature is decreased, at 2 K this  feature
signature of the energy gap is most notable, see the inset of  Fig.
\ref{fig3}. The presence of the superconducting energy gap in the
point contact spectra gives us the certitude that they are not far of
the ballistic regimen.

\begin{figure}[btp]
\begin{center}
\includegraphics[scale=1]{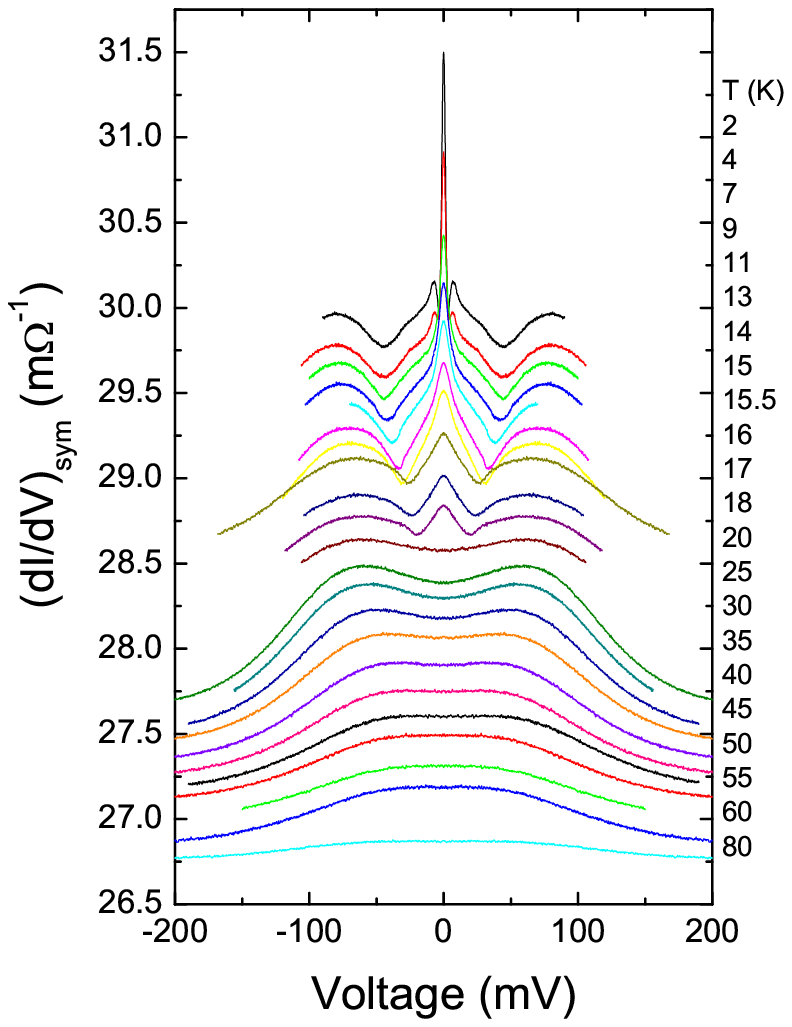}
\end{center}
\caption{(Color online) Symmetrized differential conductance as a
function of bias voltage of a Nb$_3$Sn(NS1)-Au point contact from
80 to 2 K. Note the structure below 55 K. The curves were displaced
for clarity, related to the curve measured at 2 K.}

\label{fig5}
\end{figure}

\begin{figure}[btp]
\begin{center}
\includegraphics[scale=0.5]{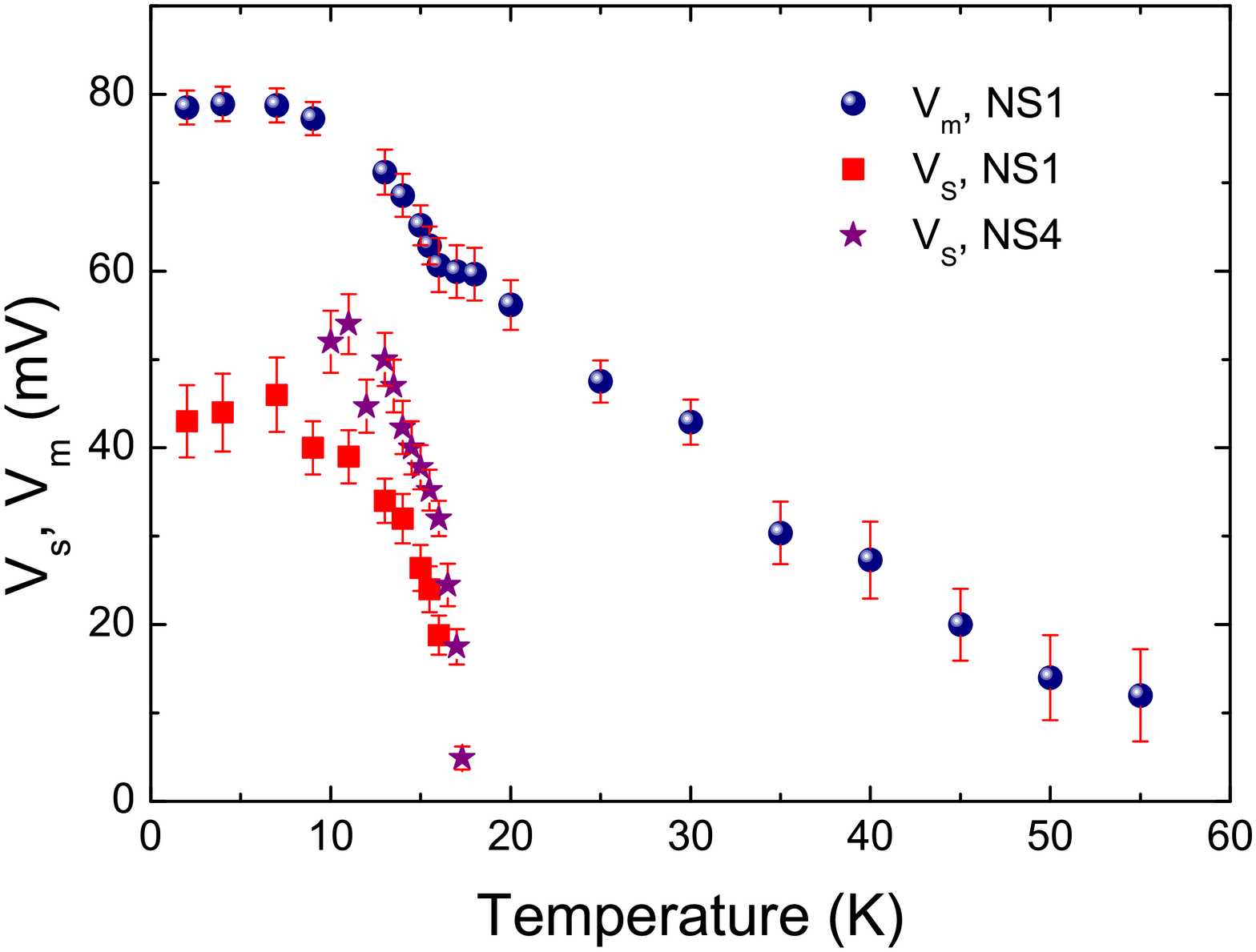}
\end{center}
\caption{(Color online) Temperature evolution of the maximum of
the differential conductance of the junction of sample NS1. Measured
from 55 to 2 K. Some interesting features can be seen: 1)
the variation of the maximum due to the high temperature anomaly,
and 2) the arising of the superconducting temperature at about 18 K.}

\label{fig6}
\end{figure}

PCS studies in those two samples show different features above the
superconducting temperature. Whereas NS1 presents structure in
$dI/dV$-V, which is related in PCS studies to the density of
electronic states, in sample NS4 there is no structure at all. A very
possible interpretation about this structure may be related to the
opening of an energy gap due to the nesting of the Fermi surface by
dimerization of Nb chains. In crystal NS1 the chains may be well
formed, in comparison to the NS4 crystal. So, NS1 shows a clear CDW
gap. Another important observation in sample NS1, on the $(dI/dV)$
characteristics is the structural asymmetry respect to zero bias. In
heterocontacts PCS studies, experts consider that this asymmetry is
produced by thermoelectric effects as a consequence of the
temperature differences in the contact region and the thermal bath.
In order to discard this contribution, the symmetric part of the
conductance was determined. The procedure consists in taking the
inverse of the symmetric differential resistance, defined as:
(dV/dI)$_{sym}$=1/2[dV/dI(V+) + dV/dI(V-)]\cite{nowack92}.

The symmetrized part, (dI/dV)$_{sym}$ as a function of V, for Fig.
\ref{fig3} is  shown in Fig. \ref{fig5}. Two maximum are in
these curves, both are at the same positive and negative
voltages. Those maxima arisen at about 45-50 K, and coincide with the
specific heat anomaly onset observed in our  measurements. Those
maxima evolve and increase with decreasing temperature. At the
superconducting transition temperature, one new feature arises; Two
symmetric minima around zero bias, that also increase as the
temperature descend. The physical meaning of this minima, as seen in
the differential conductance is not quite clear and what type of
process may be involved. However, we believe that this is the effect
of the CDW gap and superconducting gap interacting and competing for
the same parts of the Fermi surface.

The temperature variation of the maximum voltage, $V_m$, defined as
the maximum absolute value in (dI/dV)$_{sym}$ is plotted in Fig.
\ref{fig6}. The $V_m$ changes with temperature and is shown in this
figure, at temperatures from about 65 to 46 K there is not structure,
and the (dI/dV)$_{sym}$ curve shows a normal parabolic behavior (see
Fig. 5). As soon as the temperature is reduced, the $(dI/dV)$
characteristic is deformed and two maximum arise and increase. Those
maxima evolve up to the onset of the superconducting transition
temperature. At this onset about 17 K, the $(dI/dV)_{sym}$ looks almost
flat. Slightly below this temperature an additional
structure arises, which is symmetric around zero bias with two
minimum at about $\pm 20$ mV. The temperature dependence of this
minimum is plotted in Fig. \ref{fig6} for NS1 and NS4 samples. The
trend of this structure is similar in both samples, however, this
structure disappears at 10 K in NS4.

Our speculation about the physical phenomena that originate the
structure observed at high temperature in  sample NS1 by the point
contact, is that this can be attributed to nesting of parts of the
Fermi surface via the formation of a CDW gap. This CDW gap has
important consequences related to the strength of the phonon
mechanism in A15 compounds.

The Physical phenomena related to nesting 
of parts of the Fermi surface  is in agreement with the theoretical
model proposed by Bhatt and Macmillan, and also described by Bilbro and McMillan
for A15 intermetallics \cite{bhatt76}. This model is based on the Landau theory for phase transformations and on Peierls-like 
instability, from the physical viewpoint by Gor`kov \cite{gorkov73}. In the Peierls instability 
the CDW formation is due to nesting of parts of the Fermi surface, via dimerizing
chains. In A15 compounds, as in Nb$_3$Sn, chains are formed by the Nb atoms. Close to the martensitic 
transition Nb chains are dimerized. The instability decreases the ground state energy of the system,
by using a portion of electrons at the Fermi level, the immediate consequence of this phenomena is the reduction of the 
superconducting transition, in according to BCS theory.  In Fig. 6 we show our observation extracted
 from the point contact junctions, there   
the features observed are related to the modifications
 on the electronic density of states.

In Fig. 6
we observed the trend of change in $V_m$  and $V_s$, for the two samples: the voltage $V_s$ associated to the dV/dI minimum  related 
to the size of the superconducting energy  gap and the voltage $V_m$ associated to the dI/dV maximum of the CDW gap, of the sample NS1 and NS4. That voltage 
changes in different form with temperature
for the two samples. Firstly, below 20 K, $V_s$ is slightly smaller for sample NS1 than for 
 NS4, indicating  that the feature related to the superconducting energy gap is small, and 
decreases below 10 K. Also in that figure, 
the feature asociated to the CDW gap, which starts below  60 K
has a clear change at about 17 K; which is the  onset of the superconducting state. This feature follows
increasing, but now in a small trend, and below 10 K is almost constant. 
This rate of change, we assume, is due to the interaction between the superconducting state and 
the nesting of parts of Fermi surface and formation of the CDW gap. We also noted that the superconducting feature
of the NS1 sample decreases. At this point is quite interesting to  compare this behavior with   
Fig. 3 in  Bilbro and McMillan results \cite{bhatt76}.
In that figure the reduction is in the CDW gap, different to what is occurring  
in our samples, however this effect can easily  be explained considering the  
strengh of the electron phonon in Nb$_3$Sn.  This is   given by the ratio; $2\Delta/K_BT_C$, which for    
 the reported superconducting state (similar to our point contact data)
is 4.46, \cite{moore, shen, poole,
freericks02} whereas for the assumed CDW strengh this ratio is so big
as 20-26. Indicative of a CDW gap of about $2\Delta$ = 100-120 meV. 

Thus, the reduction of the martensitic anomaly will increase the
electronic population at the Fermi level, and therefore increasing
the superconducting transition. So as was metioned before, dimerizing chains reduce
 the superconducting behavior;  thus, defects or Nb deficiencies in Nb$_3$Sn or in general in A15
superconducting alloys will strengh  superconductivity.

\section{Conclusions}
Point contact spectra measured into samples NS1 and NS4 present
structure below  the martensitic anomaly. The structure is most
notable in sample NS1 which presumably has complete Nb chains, we
speculated that this structure is the feature related to the opening
of a CDW energy gap. In crystal NS4 the structure is ill-defined
because the chains are not complete dimerized, due to Nb deficiencies.

\begin{acknowledgments}
We thank S. Bernes for R-X measurements, and to F. Silvar for Helium
provisions.
\end{acknowledgments}

\end{document}